\documentclass[twocolumn,superscriptaddress,pra]{revtex4}
\usepackage{amsmath}
\usepackage{amssymb,revsymb,latexsym}

\newcommand{\nCr}[2]{\left( \begin{array}{c} #1 \\ #2 \end{array} \right)}
\newcommand{\ket}[1]{| #1 \rangle}
\newcommand{\bra}[1]{\langle #1 |}
\newcommand{\proj}[1]{\ket{#1}\!\bra{#1}}
\newcommand{\braket}[2]{\left\langle #1| #2 \right\rangle}
\newcommand{\Tr}{\operatorname{Tr}}
\newcommand{\tr}{\operatorname{Tr}}
\newcommand{\grad}{\nabla}
\newcommand{\norm}[1]{\left\| #1 \right\|_1}
\newcommand{\normtwo}[1]{\left\| #1 \right\|_2}
\newcommand{\opnorm}[1]{\left\| #1 \right\|}
\newcommand{\av}[1]{\left\langle #1 \right\rangle}
\newcommand{\rhot}{\tilde{\rho}}
\newcommand{\Omegat}{\tilde{\Omega}}

\newcommand{\identity}{\openone}

\newcommand{\eff}{\mathrm{\,eff}}
\newcommand{\interact}{\mathit{int}}
\newtheorem{lemma}{Lemma}
\newtheorem{theorem}[lemma]{Theorem}

\begin{document}

\title{The foundations of statistical mechanics from entanglement: \\Individual states vs. averages}

\author{Sandu Popescu}
\affiliation{H.H.Wills Physics Laboratory, University of Bristol, %
Tyndall Avenue, Bristol BS8 1TL, U.K.}
\affiliation{Hewlett-Packard Laboratories, Stoke Gifford, Bristol BS12 6QZ, U.K.}

\author{Anthony J. Short}
\affiliation{H.H.Wills Physics Laboratory, University of Bristol, %
Tyndall Avenue, Bristol BS8 1TL, U.K.}

\author{Andreas Winter}
\affiliation{Department of Mathematics, University of Bristol, %
University Walk, Bristol BS8 1TW, U.K.}

\begin{abstract}
We consider an alternative approach to the foundations of
statistical mechanics, in which subjective randomness,
ensemble-averaging or time-averaging are not required. Instead,
the universe (i.e. the system together with a sufficiently large
environment) is in a quantum pure state subject to a global
constraint, and thermalisation results from  entanglement between
system and environment. We formulate and prove a ``General
Canonical Principle'', which states that the system will be
thermalised for almost all pure states of the universe, and
provide rigorous quantitative bounds using Levy's Lemma.
\end{abstract}

\maketitle

\section{Introduction}
\label{intro}

Despite many years of research, the foundations of statistical
mechanics remain a controversial subject. Crucial questions
regarding the role of probabilities and entropy (which are viewed
both as measures of ignorance and objective properties of the
state) are not satisfactorily resolved, and the relevance of time
averages and ensemble averages to individual physical systems is
unclear.

Here we adopt a fundamentally new viewpoint suggested by Yakir
Aharonov \cite{aharonov}, which is uniquely quantum, and which
does not rely on any ignorance probabilities in the description of
the state. We consider the global state of a large isolated
system, the `universe', to be a quantum \emph{pure} state. Hence
there is no lack of knowledge about the state of the universe, and
the entropy of the universe is zero. However, when we consider
only part of the universe (that we call the `system'), it is
possible that its state will not be pure, due to quantum
entanglement with the rest of the universe (that we call the
`environment'). Hence there is an \emph{objective} `lack of
knowledge' about the state of the system, even though we know
everything about the state of the universe. In such cases, the
entropy of the system is non-zero, even though we have introduced
no randomness and the universe itself has zero entropy.

Furthermore, interactions between the system and environment can
objectively increase both the entropy of the system and that of
the environment by increasing their entanglement. It is
conceivable that this is the mechanism behind the second law of
thermodynamics. Indeed, as information about the system will tend
to leak into (and spread out in) the environment, we might well
expect that their entanglement (and hence entropy) will increase
over time in accordance with the second law.

The above ideas provide a compelling vision of the foundations of
statistical mechanics. Such a viewpoint has been independently
proposed recently by Gemmer et al. \cite{mahler}.

In this paper, we address one particular aspect of the above
programme. We show that thermalisation is a \emph{generic}
property of pure states of the universe, in the sense that for
almost all of them, the reduced state of the system is  the
\emph{canonical} mixed state. That is, not only is the state of
the system mixed (due to entanglement with the rest of the
universe), but it is in precisely the state we would expect from
standard statistical arguments.

In fact, we prove a stronger result. In the standard statistical
setting, energy constraints are imposed on the state of the
universe, which then determine a corresponding temperature and
canonical state for the system. Here we consider that states of
the universe are subject to arbitrary constraints. We then show
that almost every pure state of the universe subject to those
constraints is such that the system is in the corresponding
generalised canonical state.

Our results are kinematic, rather than dynamical. That is, we do
not consider any particular unitary evolution of the global state,
and we do not show that thermalisation of the system occurs.
However, because almost all states of the universe are such that
the system is in a canonical thermal state, we anticipate that
most evolutions will quickly carry a state in which the system is
not thermalised to one in which it is, and that the system will
remain thermalised for most of its evolution.

 A key ingredient in our analysis is Levy's Lemma
\cite{levy1, levy2}, which plays a similar role to the law of
large numbers and governs the properties of typical states in
large-dimensional Hilbert spaces. Levy's Lemma
 has already been used in quantum information theory to study
entanglement and other correlation properties of random states in
large bipartite systems \cite{hayden}. It provides a very powerful
tool with which to evaluate functions of randomly chosen quantum
states.

The structure of this paper is as follows. In section II we
present our main result in the form of a General Canonical
Principle. In section III we support this principle with precise
mathematical theorems. In section IV we introduce Levy's Lemma,
which is used in sections V and VI to provide proofs of our main
theorems. Section VII illustrates these results with the simple
example of spins in a magnetic field. Finally, in section VIII we
present our conclusions.

\section{General Canonical Principle}
\label{sec:principles}

Consider a large quantum mechanical system, `the universe', that
we decompose into two parts, the `system' $S$ and the
`environment' $E$. We will assume that the dimension of the
environment is much larger than that of the system. Consider now
that the state of the universe obeys some global constraint $R$.
We can represent this quantum mechanically by restricting the
allowed states of the system and environment to a subspace
$\mathcal{H}_R$ of the total Hilbert space:
\begin{equation}
\mathcal{H}_R \subseteq \mathcal{H}_S \otimes \mathcal{H}_E,
\end{equation}
where $\mathcal{H}_S$ and $\mathcal{H}_E$ are the Hilbert spaces
of the system and environment, with dimensions $d_S$ and $d_E$
respectively. In standard statistical mechanics $R$ would
typically be a restriction on the total energy of the universe,
but here we leave $R$ completely general.

We define $\mathcal{E}_R$, the \textbf{equiprobable state of the
universe corresponding to the restriction $R$}, by
\begin{equation}
\mathcal{E}_R = \frac{\identity_R}{d_R},
\end{equation}
where $\identity_R$ is the identity (projection) operator on
$\mathcal{H}_R$, and $d_R$ is the dimension of $\mathcal{H}_R$.
$\mathcal{E}_R$ is the maximally mixed state in $\mathcal{H}_R$,
in which each pure state has equal probability. This corresponds
to the standard intuition of assigning equal a priori
probabilities to all states of the universe consistent with the
constraints.

We define $\Omega_S$, the \textbf{canonical state of the system
corresponding to the restriction $R$}, as the quantum state of the
system when the universe is in the equiprobable state
$\mathcal{E}_R$. The canonical state of the system $\Omega_S$ is
therefore obtained by tracing out the environment in the
equiprobable state of the universe:
\begin{equation}
\Omega_S = \Tr_E ( \mathcal{E}_R ). \label{canonical_eqn}
\end{equation}

We now come to the main idea behind our paper.

As described in the introduction, we now consider that the
universe is in a pure state $\phi$, and \emph{not} in the mixed
state $\mathcal{E}_R$ (which represents a subjective lack of
knowledge about its state). We prove that despite this, the state
of the system is very close to the canonical state $\Omega_S$ in
almost all cases. That is, for almost every pure state of the
universe, the system behaves as if the universe were actually in
the equiprobable mixed state $\mathcal{E}_R$.

We now state this basic qualitative result as a general principle,
that will subsequently be refined by quantitative theorems:

\medskip

\textbf{General Canonical Principle:} \emph{Given a sufficiently
small subsystem of the universe, almost every pure state of the
universe is such that the subsystem is approximately in the
canonical state $\Omega_S$.}

\medskip

Recalling that the canonical state of the system $\Omega_S$ is, by
definition, the state of the system when the universe is in the
equiprobable state $\mathcal{E}_R$ we can interpret the above
principle as follows:

\medskip

\textbf{Principle of Apparently Equal a priori Probability:}
\emph{For almost every pure state of the universe, the state of a
sufficiently small subsystem is approximately the same as if the
universe were in the equiprobable state $\mathcal{E}_R$. In other
words, almost every pure state of the universe is locally (i.e. on
the system) indistinguishable from $\mathcal{E}_R$.}

\medskip

For an arbitrary pure state $\ket{\phi}$ of the universe, the
state of the system alone is given by
\begin{equation}
\rho_S = \Tr_E(\proj{\phi}).
\end{equation}
Our principle states that for almost all states $\ket{\phi} \in
\mathcal{H}_R$,
\begin{equation}
\rho_S \approx \Omega_S.
\end{equation}

Obviously, the above principle is stated qualitatively. To express
these results quantitatively, we need to carefully define what we
mean by a sufficiently small subsystem, under what distance
measure $\rho_S \approx \Omega_S$, and how good this approximation
is. This will be done in the remaining sections of the paper.

We emphasise that the above is a generalised principle, in the
sense that the restriction $R$ imposed on the states of the
universe is completely arbitrary (and is not necessarily the usual
constraint on energy or other conserved quantities). Similarly,
the canonical state $\Omega_S$ is not necessarily the usual
thermal canonical state, but is defined relative to the arbitrary
restriction $R$ by equation (\ref{canonical_eqn}).

To connect the above principle to standard statistical mechanics,
all we have to do is to consider the restriction $R$ to be that
the total energy of the universe is close to $E$, which then sets
the temperature scale $T$. The total Hamiltonian of the universe
$H_U$ is given by
\begin{equation}
H_U = H_S + H_E + H_{\interact},
\end{equation}
where $H_S$ and $H_E$ are the Hamiltonians of the system and
environment respectively, and $H_{int}$ is the interaction
Hamiltonian between the system and environment. In the standard
situation, in which $H_{int}$ is small and the energy spectrum of
the environment is sufficiently dense and uniform, the canonical
state $\Omega_S^{(E)}$ can be computed using standard techniques,
and shown to be
\begin{equation}
\Omega_S^{(E)} \propto \exp\left(-\frac{H_S}{\mathrm{k}_B
T}\right).
\end{equation}

This allows us to state the thermal canonical principle that
establishes the validity (at least kinematically) of the viewpoint
expressed in the introduction.

\bigskip \noindent \textbf{Thermal Canonical Principle:}
\emph{Given that the total energy of the universe is approximately
$E$, interactions between the system and the rest of the universe
are weak, and that the energy spectrum of the universe is
sufficiently dense and uniform, almost every pure state of the
universe is such that the state of the system alone is
approximately equal to the thermal canonical state
$e^{-\frac{H_S}{\mathrm{k}_B T}}$, with temperature $T$
(corresponding to the energy $E$)} \bigskip

We emphasise here that our contribution in this paper is to show
that $\rho_S \approx \Omega_S$, and has nothing to do with showing
that $\Omega_S \propto e^{-\frac{H_S}{\mathrm{k}_B T}}$, which is
a standard problem in statistical mechanics.

Finally, we note that the General Canonical Principle applies also
in the case where the interaction between the system and
environment is not small. In such situations, the canonical state
of the system is no longer $e^{-\frac{H_S}{\mathrm{k}_B T}}$,
since the behaviour of the system will depend very strongly on
$H_{int}$. Nevertheless, the general principle remains valid for
the corresponding generalised canonical state $\Omega_S$.
Furthermore our principle will apply to arbitrary restrictions $R$
that have nothing to do with energy, which may lead to many
interesting insights.

\section{Quantitative setup \protect\\ and main theorems} \label{theorem_sec}

We now formulate and prove precise mathematical theorems
correponding to the General Canonical Principle stated in the
previous section.

As a measure of the distance between $\rho_S$ and $\Omega_S$, we
use the trace-norm $\norm{ \rho_S - \Omega_S}$, where
\cite{trace-dist}
\begin{equation}
  \norm{M} = \Tr |M| = \Tr \sqrt{M^{\dag} M},
\end{equation}
as this distance will be small if and only if it would be hard for
any measurement to tell $\rho_S$ and $\Omega_S$ apart. Indeed,
$\norm{M} = \sup_{\|O\|\leq 1} \tr(MO)$, where the maximisation is
over all operators (observables) $O$ with operator norm bounded by
$1$.

In our analysis, we also make use of the Hilbert-Schmidt norm
$\normtwo{M} = \sqrt{\Tr(M^{\dag} M)}$, which is easier to
manipulate than $\norm{M}$. However, we only use this for
intermediate calculational purposes, as it does not have the
desirable physical properties of the trace-norm. In particular
$\normtwo{ \rho_S - \Omega_S}$ can be small even when the two
states are orthogonal for high-dimensional systems.

Throughout this paper we denote by $\av{\cdot}$ the average over
states $\ket{\phi} \in \mathcal{H}_R$ according to the uniform
distribution. For example, it is easy to see that $\Omega_S =
\av{\rho_S}$.

We will prove the following theorems:
\begin{theorem}
  \label{thm:main}
  For a randomly chosen state $\ket{\phi} \in \mathcal{H}_R
  \subseteq \mathcal{H}_S \otimes \mathcal{H}_E$ and arbitrary $\epsilon>0$,
  the distance between the reduced density matrix of the system $\rho_S=\Tr(\proj{\phi})$ and the
  canonical state $\Omega_S=\Tr \mathcal{E}_R$ is given probabilistically by
  \begin{equation}
    \mathrm{Prob}\big[ \norm{\rho_S-\Omega_S} \geq \eta \big] \leq \eta',
  \end{equation}
  where
  \begin{eqnarray}
    \eta  &=& \epsilon  + \sqrt{ \frac{d_S}{d_E^{\eff}}}, \\
    \eta' &=& 2 \exp \left( - C d_R \epsilon^2 \right).
  \end{eqnarray}
  In these expressions, $C$ is a positive constant (given by $C=(18
  \pi^3)^{-1}$), $d_S$ and $d_R$ are the dimensions of
  $\mathcal{H}_S$ and $\mathcal{H}_R$ respectively, and $d_E^{\eff}$
  is a measure of the effective size of the environment, given by
  \begin{equation} \label{deff_eqn}
    d_E^{\eff} = \frac{1}{\Tr \Omega_E^2} \geq \frac{d_R}{d_S},
  \end{equation}
  where $\Omega_E = \tr_S \mathcal{E}_R$.
  Both $\eta$ and $\eta'$ will be small quantities, and thus the
  state will be close to the canonical state with high probability,
  whenever $d_E^{\eff} \gg d_S$ (i.e. the effective dimension of the
  environment is much larger than that of the system) and $d_R \epsilon^2 \gg
  1 \gg \epsilon$. This latter condition can be ensured when $d_R \gg 1$
  (i.e. the total accessible space is large), by choosing $\epsilon = d_R^{-1/3}$.
\end{theorem}

This theorem gives rigorous meaning to our statements in section
\ref{sec:principles} about thermalisation being achieved for
`almost all' states: we have an exponentially small bound on the
relative volume of the exceptional set, i.e. on the probability of
finding the system in a state that is far from the canonical
state. Interestingly, the exponent scales with the dimension of
the space ${\cal H}_R$ of the constraint, while the deviation from
the canonical state is characterised by the ratio between the
system size and the effective size of the environment, which makes
intuitive sense.

Theorem \ref{thm:main} provides a bound on the distance between
$\rho_S$ and $\Omega_S$, but in many situations we can further
improve it. Often the system does not really occupy all of its
Hilbert space $\mathcal{H}_S$, and also the estimate of the
effective environment dimension $d_E^{\eff}$ may be too small, due
to exceptionally large eigenvalues of $\Omega_E = \tr_S
\mathcal{E}_R$. By cutting out these non-typical components
(similar to the well-known method of projecting onto the typical
subspace), we can optimize the bound obtained, as we will show in
Theorem \ref{thm:main-approximate}. The benefits of this
optimization will be apparent in section \ref{spin_sec}, where we
consider a particular example.

\begin{theorem}
  \label{thm:main-approximate}
  Assume that there exists some bounded positive operator $X_R$ on $\mathcal{H_R}$ satisfying
$0 \leq X_R \leq \identity$ such that, with $\tilde{\mathcal{E}}_R
= \sqrt{X_R} \mathcal{E}_R \sqrt{X_R}$,
  \begin{equation}
  \Tr(\tilde{\mathcal{E}}_R) = \Tr \bigl( \mathcal{E}_R X_R \bigr)
                      \geq 1-\delta.
  \end{equation}
  (I.e. the probability of obtaining the outcome corresponding to measurement operator $X_R$ in
  a generalised measurement (POVM) on $\mathcal{E}_R$ is approximately one.)

  Then, for a randomly chosen state $\ket{\phi} \in \mathcal{H}_R
  \subseteq \mathcal{H}_S \otimes \mathcal{H}_E$ and $\epsilon>0$,
  \begin{equation}
    \mathrm{Prob}\big[ \norm{\rho_S-\Omega_S} \geq \tilde\eta \big] \leq \tilde\eta',
  \end{equation}
  where
  \begin{eqnarray}
    \tilde\eta  &=&  \epsilon + \sqrt{ \frac{\tilde{d}_S}{\tilde{d}_E^{\eff}}}
                        + 4 \sqrt{\delta}, \\
    \tilde\eta' &=& 2 \exp \left( - C  d_R \epsilon^2\right).
  \end{eqnarray}
  Here, $C$ and $d_R$ are as in Theorem~\ref{thm:main},
  $\tilde{d}_S$ is the dimension of the support of $X_R$ in $\mathcal{H}_S$,
  and $\tilde{d}_E^{\eff}$ is the effective size
  of the environment after applying $X_R$, given by
  \begin{equation}
    \tilde{d}_E^{\eff} = \frac{1}{\Tr \Omegat_E^2} \geq \frac{d_R}{\tilde{d}_S},
  \end{equation}
  where $\Omegat_E = \tr_S (\tilde{\mathcal{E}}_R)$. In many
  situations $\delta$ can be made very small, while at the same
  time improving the relation between system and effective environment
  dimension. Note that the above is essentially the technique
  of the smooth (quantum) R\'{e}nyi entropies \cite{smooth-renyi, hayden:winter}:
  $\log \tilde{d}_S$ is related to $S_0^\delta(\Omega_S)$ and $\log \tilde{d}_E^{\eff}$ to
  $S_2^\delta(\Omega_E)$.
\end{theorem}

In the process of proving these theorems, we also obtain the
following subsidiary results:

\begin{enumerate}
\item The average distance between the system's reduced density matrix for
  a randomly chosen state and the canonical state will be small
  whenever the effective environment size is larger than the system.
  Specifically,
  \begin{equation}
    \av{\norm{\rho_S - \Omega_S}} \leq \sqrt{\frac{d_S}{d_E^{\eff}}}
       \leq \sqrt{\frac{d_S^2}{d_R}},
  \end{equation}
  where the effective dimension of the environment $d_E^{\eff}$
  is given by (\ref{deff_eqn}).

\item With high probability, the expectation value of a bounded
  observable $O_S$ on the system for a randomly chosen state will be
  very similar to its expectation value in the canonical state
  whenever $d_R \gg 1$. Specifically,
  \begin{equation}\begin{split}
    {\rm Prob}&\big[ \left|\Tr(O_S \rho_S) - \Tr(O_S \Omega_S) \right|
                                                     \geq d_R^{-1/3} \big] \\
              &\phantom{=========}
               \leq 2 \exp\left( - \frac{C d_R^{1/3}}{\opnorm{O_S}^2} \right),
  \end{split}\end{equation}
  where $C$ is a constant.
\end{enumerate}

In our analysis we use two alternative methods, with the hope that
the different mathematical techniques employed will aid in future
exploration of the field.

\section{Levy's Lemma}
\label{sec:levy} A major component in the proofs of the following
sections is the mathematical theorem known as Levy's Lemma
\cite{levy1,levy2}, which states that when a point $\phi$ is
selected at random from a hypersphere of high dimension and
$f(\phi)$ does not vary too rapidly, then $f(\phi) \approx \av{f}$
with high probability:

\begin{lemma}
(\textbf{Levy's Lemma}) Given a function $f: \mathbb{S}^{d}
\rightarrow \mathbb{R}$ defined on the $d$-dimensional hypersphere
$\mathbb{S}^{d}$, and a point $\phi \in \mathbb{S}^{d}$ chosen
uniformly at random,
\begin{equation} \label{levy_eqn}
{\rm Prob} \big[ \left|f(\phi) - \langle f \rangle \right| \geq
\epsilon \big] \leq 2 \exp \left( \frac{- 2 C (d+1) \epsilon^2 }{
\eta^2 } \right)
\end{equation}
where $\eta$ is the Lipschitz constant of $f$, given by $\eta =
\sup | \grad f|$, and $C$ is a positive constant (which can be
taken to be $C=(18 \pi^3)^{-1}$).
\end{lemma}
Due to normalisation, pure states in $\mathcal{H}_R$ can be
represented by points on the surface of a $(2 d_R-1)$-dimensional
hypersphere $\mathbb{S}^{2 d_R-1}$, and hence we can apply Levy's
Lemma to functions of the randomly selected quantum state $\phi$
by setting $d=2 d_R-1$. For such a randomly chosen state
$\ket{\phi} \in \mathcal{H}_R$, we wish to show that $\norm{\rho_S
- \Omega_S}\approx 0$ with high probability.

\section{Method I: Applying Levy's \protect\\ Lemma to $\norm{\rho_S-\Omega_S}$}
\label{sec:use-levy-directly}
In this section, we consider the consequences of applying Levy's
Lemma directly to the distance between
$\rho_S=\Tr_E(\proj{\phi})$ and $\Omega_S$, by choosing
\begin{equation}
f(\phi)=\norm{\rho_S-\Omega_S}.
\end{equation}
in (\ref{levy_eqn}). As we prove in appendix \ref{lipschitz_app},
the function $f(\phi)$ has Lipschitz constant $\eta \leq 2$.
Applying Levy's Lemma to $f(\phi)$ then gives:
\begin{equation} \label{init_levy_eqn}
\mathrm{Prob}\Big[ \,\Big| \norm{\rho_S-\Omega_S} -
\av{\norm{\rho_S-\Omega_S}} \Big| \geq \epsilon \Big] \leq 2 e^{-
C d_R \epsilon^2  }.
\end{equation}
To obtain Theorem 1, we rearrange this equation to get
\begin{equation} \label{pre_calc_eqn}
\mathrm{Prob}\big[ \norm{\rho_S-\Omega_S} \geq \eta \big] \leq
\eta'
\end{equation}
where
\begin{eqnarray}
\eta &=& \epsilon + \av{\norm{\rho_S-\Omega_S}} \label{pre_calc_eqn2} \\
\eta' &=& 2 \exp \left( - C d_R \epsilon^2 \right).
\end{eqnarray}

The focus of the following subsections is to obtain a bound on
$\av{\norm{\rho_S-\Omega_S}}$. In section \ref{av_norm_sec} we
show that
\begin{equation}\label{av_norm_eq}
\av{\norm{\rho_S - \Omega_S}} \leq \sqrt{\frac{d_S}{d_E^{\eff}}}
\end{equation}
where $d_E^{\eff}$ is a measure of the effective size of the
environment, given by (\ref{deff_eqn}). Inserting equation
(\ref{av_norm_eq}) in (\ref{pre_calc_eqn2}) we obtain Theorem 1.

Typically $d_R \gg 1$ (the total number of accessible states is
large) and hence by choosing $\epsilon=d_R^{-1/3}$ we can ensure
that both $\epsilon$ and $\eta'$ are small quantities. When it is
also true that $d_E^{\eff} \gg d_S$ (the environment is much
larger than that of the system) both $\eta$ and $\eta'$ will be
small quantities, leading to $\norm{\rho_S - \Omega_S} \approx 0$
with high probability.

To obtain Theorem 2, we consider a generalised measurement which
has an almost certain outcome for the equiprobable state
$\mathcal{E}_R \in \mathcal{H_R}$, and apply the corresponding
measurement operator before proceeding with our analysis. By an
appropriate choice of measurement operator, the ratio of the
system and environment's effective dimensions can be significantly
improved (as shown by the example in section
\ref{typical_sub_sec}).

\subsection{Calculating $\av{\norm{\rho_S-\Omega_S}}$ }
As mentioned in section \ref{theorem_sec}, although
$\norm{\rho_S-\Omega_S}$ is a physically meaningful quantity, it
is difficult to work with directly, so we first relate it to the
Hilbert-Schmidt norm $\normtwo{\rho_S - \Omega_S}$. The two norms
are related by
\begin{equation}
\norm{\rho_S-\Omega_S} \leq \sqrt{d_S} \normtwo{\rho_S-\Omega_S},
\end{equation}
as proved in Appendix \ref{norm_app}.

Expanding $\normtwo{\rho_S - \Omega_S}$ we obtain
\begin{eqnarray}
  \av{\normtwo{\rho_S - \Omega_S}}
     &\leq& \sqrt{ \av{\normtwo{\rho_S - \Omega_S}^2}} \\
     &=&    \sqrt{\av{\Tr(\rho_S - \Omega_S)^2}} \nonumber \\
     &=&    \sqrt{\av{\Tr \rho_S^2} - 2 \Tr(\av{\rho_S} \Omega_S)
                                    + \Tr \Omega_S^2  } \nonumber \\
     &=&    \sqrt{ \av{\Tr \rho_S^2} - \Tr \Omega_S^2 },
\end{eqnarray}
and hence
\begin{equation} \label{expanded_norm_eqn}
  \av{\norm{\rho_S - \Omega_S}}
            \leq \sqrt{d_S \left(\av{\Tr \rho_S^2} - \Tr \Omega_S^2 \right)}
\end{equation}

\subsection{Calculating $\av{\Tr(\rho_S^2)}$}
In this section we show the fundamental inequality
\begin{equation}
\av{\Tr \rho_S^2} \leq \Tr \av{\rho_S}^2 + \Tr \av{\rho_E}^2.
\end{equation}
The following calculations
and estimates are closely related to the arguments used in random
quantum channel coding \cite{lloyd-shor} and random entanglement
distillation (see \cite{HOW}).

To calculate $\av{\Tr \rho_S^2}$, it is helpful to introduce a
second copy of the original Hilbert space, extending the problem
from $\mathcal{H}_R$ to $\mathcal{H}_R \otimes \mathcal{H}_{R'}$
where $\mathcal{H}_{R'} \subseteq \mathcal{H}_{S'} \otimes \mathcal{H}_{E'}$.

Note that
\begin{eqnarray}
  \Tr_S \rho_S^2 &=& \sum_k (\rho_{kk})^2 \nonumber \\
                 &=& \sum_{k,l,k',l'} (\rho_{kl}) (\rho_{k'l'})
                                      \braket{kk'}{ll'}\braket{l'l}{kk'}\nonumber \\
                 &=& \Tr_{SS'}\bigl( (\rho_S \otimes \rho_{S'}) F_{SS'} \bigr),
                                                                  \label{flip_eqn}
\end{eqnarray}
where $F_{SS'}$ is the flip (or swap) operation $S\leftrightarrow S'$:
\begin{equation}
  F_{SS'} = \sum_{S,S'} \ket{s'}\!\bra{s}_S \otimes \ket{s}\!\bra{s'}_{S'},
\end{equation}
and hence
\begin{equation}
  \Tr_S \rho_S^2 = \Tr_{RR'}\bigl( (\proj{\phi} \!\otimes\! \proj{\phi})_{RR'}
                                      (F_{SS'}\otimes \identity_{EE'}) \bigr).
\end{equation}
So, our problem reduces to the calculation of
\begin{equation}
  V \equiv \av{\proj{\phi} \otimes \proj{\phi}}= \int \proj{\phi}
\otimes \proj{\phi} \,\textrm{d}\phi.
\end{equation}
As $V$ is invariant under operations of the form $V \rightarrow (U
\otimes U) V (U^{\dag} \otimes U^{\dag}) $ for any unitary $U$,
representation theory implies that
\begin{equation}
V=\alpha \Pi^{\textrm{sym}}_{RR'} + \beta
\Pi^{\textrm{anti}}_{RR'},
\end{equation}
where $\Pi^{\textrm{sym/anti}}_{RR'}$ are projectors onto the
symmetric and antisymmetric subspaces of $\mathcal{H}_R \otimes
\mathcal{H}_{R'}$ respectively, and $\alpha$ and $\beta$ are
constants. As
\begin{equation}
(\proj{\phi} \otimes \proj{\phi}) \frac{1}{\sqrt{2}} (\ket{ab} -
\ket{ba}) = 0 \quad \forall \, a, b,\phi,
\end{equation}
it is clear that $\beta=0$, and as $V$ is a normalised state,
\begin{equation}
\alpha = \frac{1}{\textrm{dim}(RR'_{\textrm{sym}})} =
\frac{2}{d_R(d_R+1)}.
\end{equation}
Hence
\begin{equation}
  \av{\proj{\phi} \otimes \proj{\phi}} = \frac{2}{d_R(d_R+1)}
                                         \Pi^{\textrm{sym}}_{RR'}.
\end{equation}
and therefore
\begin{equation}
  \av{\Tr_S\rho_S^2} =
                        \Tr_{RR'}\left( \left( \frac{ 2\,
                        \Pi^{\textrm{sym}}_{RR'}}{d_R(d_R+1)}\right)
                                        (F_{SS'}\otimes \identity_{EE'}) \right).
\end{equation}
To proceed further we perform the substitution
\begin{equation}
  \Pi^{\textrm{sym}}_{RR'} = \frac{1}{2} \left( \identity_{RR'} + F_{RR'} \right),
\end{equation}
where $F_{RR'}$ is the flip operator taking $R \leftrightarrow
R'$. Noting that $F_{RR'}=\identity_{RR'}(F_{SS'}\otimes
F_{EE'})$, this gives
\begin{eqnarray}
  \av{\Tr_S \rho_S^2} &=&
                          \Tr_{RR'}\left( \left(
                          \frac{\identity_{RR'}}{d_R(d_R+1)}
                          \right) (F_{SS'}\otimes \identity_{EE'})\right)\nonumber\\
                      & &\quad\!\!
                         +
                          \Tr_{RR'}\left( \left(
                          \frac{\identity_{RR'}}{d_R(d_R+1)}
                          \right) (\identity_{SS'}\otimes F_{EE'})\right)\nonumber\\
                      &\leq&  \Tr_{RR'}\left( \left(\frac{\identity_R}{d_R} \otimes
                                \frac{\identity_{R'}}{d_R}\right) (F_{SS'}\otimes
                                  \identity_{EE'})\right) \nonumber\\
                      & &\quad\!\!
                         + \Tr_{RR'}\left( \left(\frac{\identity_R}{d_R} \otimes
                             \frac{\identity_{R'}}{d_R}\right) (\identity_{SS'}\otimes
                               F_{EE'})\right) \nonumber\\
                      &=& \Tr_{SS'}\bigl( (\Omega_S \otimes \Omega_S)
                                                             F_{SS'} \bigr)\nonumber \\
                      & &\quad\!\! + \Tr_{EE'}\bigl( (\Omega_E \otimes \Omega_E)
                                                             F_{EE'} \bigr).
\end{eqnarray}
Hence from equation (\ref{flip_eqn}),
\begin{equation}\begin{split}
  \av{\Tr_S(\rho_S^2)} &\leq \Tr_S \Omega_S^2 + \Tr_E \Omega_E^2 \\
                       &=    \Tr \av{\rho_S}^2 + \Tr \av{\rho_E}^2
\end{split}\end{equation}

\subsection{Bounding $\av{\norm{\rho_S - \Omega_S}}$}
\label{av_norm_sec}
Inserting the results of the last section in equation
(\ref{expanded_norm_eqn}) we obtain
\begin{equation}
\av{\norm{\rho_S - \Omega_S}} \leq \sqrt{d_S \Tr_E \Omega_E^2}
\end{equation}

Intuitively, we can understand this equation by defining
\begin{equation}
d_E^{\eff} = \frac{1}{\Tr_E \Omega_E^2 },
\end{equation}
as the effective dimension of the environment in the canonical
state. If all of the non-zero eigenvalues of $\Omega_E$ were of
equal weight this would simply correspond to the dimension of
$\Omega_E$'s support, but more generally it will measure the
dimension of the space in which the environment is most likely to
be found. When there is no constraint on the accessible states of
the environment, such that $\mathcal{H_R} = \mathcal{H}'_S \otimes
\mathcal{H}_E$ then $d_E^{\eff}= d_E$

Denoting the eigenvalues of $\Omega_E$ by $\lambda_E^k$ (with
maximum eigenvalue $\lambda_E^{max}$), it is also interesting to
note that
\begin{eqnarray}
\Tr_E \Omega_E^2  &=& \sum_k (\lambda_E^k)^2 \nonumber \\ &\leq&
\lambda_E^{\max} \sum_k \lambda_E^k \nonumber \\ &=&
\max_{\ket{\psi_E}}\, \bra{\psi_E}
\Tr_S\left(\frac{\identity_R}{d_R}\right) \ket{\psi_E} \nonumber \\
&=& \max_{\ket{\psi_E}}\, \sum_s \bra{s\, \psi_E }
\frac{\identity_R}{d_R} \ket{s\, \psi_E } \nonumber \\ &\leq&
\frac{d_S}{d_R}. \label{dEeff_eqn}
\end{eqnarray}
Hence $d_E^{\eff} \geq d_R/d_S$, and we obtain the final result
that
\begin{equation} \label{av_norm_eqn}
\av{\norm{\rho_S - \Omega_S}} \leq \sqrt{\frac{d_S}{d_E^{\eff}}}
\leq \sqrt{\frac{d_S^2}{d_R}}.
\end{equation}
The average distance $\av{\norm{\rho_S - \Omega_S}}$ will
therefore be small whenever the effective size of the environment
is much larger than that of the system ($d_E^{\eff} \gg d_S$).

Inserting the results of equation (\ref{av_norm_eqn}) into
equation (\ref{pre_calc_eqn2}) gives Theorem 1.

\subsection{Improved bounds using restricted subspaces}

As mentioned in section \ref{theorem_sec}, in many cases it is
possible to improve the bounds obtained from Theorem
\ref{thm:main} by projecting the states onto a typical subspace
before proceeding with the analysis. This can allow one to
decrease the effective dimension of the system $d_S$ (by
eliminating components with negligible amplitude), and increase
the effective dimension of the environment $d^{\eff}_E=(\Tr
\Omega_E^2)^{-1}$ (by eliminating components of $\Omega_E$ with
disproportionately high amplitudes), whilst leaving the
equiprobable state $\mathcal{E}_R$ largely unchanged.

To allow for the most general possibility, we consider a
generalised measurement operator $X_R$ satisfying $0 \leq X_R \leq
\identity$ (of which a projector is a special case), which has
high probability of being satisfied by $\mathcal{E}_R$, such that
\begin{equation} \label{delta_eqn}
\Tr_R(\mathcal{E}_R X_R) \geq 1-\delta.
\end{equation}

We denote the dimension of the support of $X_R$ in $\mathcal{H}_S$
by $\tilde{d}_S$, which will play the role of $d_S$ in the revised
analysis \cite{support}. The bounds on $\av{\norm{\rho_S -
\Omega_S}}$ will be optimized by choosing $X_R$ such that
$\tilde{d}_S$ is as small as possible, and $\tilde{d}_E^{\eff}$ as
large as possible.

We also define the sub-normalised states obtained after
measurement of $X_R$:
\begin{eqnarray}
\ket{\tilde{\phi}} &=& \sqrt{X_R} \ket{\phi} \\
\tilde{\mathcal{E}}_R &=&  \sqrt{X_R}\, \mathcal{E}_R\, \sqrt{X_R} = \frac{X_R}{d_R} \\
\Omegat_S &=& \Tr_E(\tilde{\mathcal{E}}_R) \\
\Omegat_E &=& \Tr_S(\tilde{\mathcal{E}}_R)
\end{eqnarray}

Applying the same analysis as in the previous sections to these
states, we find
\begin{eqnarray}
  \av{\Tr_S \rhot_S^2} &=& \Tr_{RR'}\bigl( \av{\proj{\tilde{\phi}} \!\otimes\!
                             \proj{\tilde{\phi}}}(F_{SS'}\otimes \identity_{EE'})
                            \bigr) \nonumber \\
                        &=& \Tr_{RR'}\Biggl(
                          \frac{(X_R \otimes X_R)\Pi^{sym}_{RR'}}{d_R(d_R+1)}  (F_{SS'}\otimes \identity_{EE'})\Biggr) \nonumber \\
                        &\leq&  \Tr_{RR'}\left( \left(\frac{X_R}{d_R} \otimes
                                \frac{X_{R'}}{d_R}\right) (F_{SS'}\otimes
                                  \identity_{EE'})\right) \nonumber \\
                      & &\quad\!
                         + \Tr_{RR'}\!\left( \left(\frac{X_R}{d_R} \otimes
                             \frac{X_{R'}}{d_R}\right) (\identity_{SS'}\otimes
                               F_{EE'})\right) \nonumber \\
                      &=& \Tr_S \Omegat_S^2 + \Tr_E \Omegat_E^2,
\end{eqnarray}
where in the second equality we have used the fact that
$[\sqrt{X_R} \otimes \sqrt{X_{R'}},\, \Pi^{sym}_{RR'}] =0$. From
the analogue of equation (\ref{expanded_norm_eqn}) we can then
obtain
\begin{equation}
\av{\norm{\rhot_S - \Omegat_S}} \leq
\sqrt{\frac{\tilde{d}_S}{\tilde{d}_E^{\eff}}},
\end{equation}
where (using the analogue of (\ref{dEeff_eqn}))
\begin{equation}
\tilde{d}^{\eff}_{E} = \frac{1}{\Tr_E \Omegat_E^2} \geq
\frac{d_R}{\tilde{d}_S}.
\end{equation}

To transform this bound on $\norm{\rhot_S - \Omegat_S}$ into a
bound on $\norm{\rho_S - \Omega_S}$, we note that
\begin{equation} \label{av_sum_eqn}
\norm{\rho_S - \Omega_S} \leq  \norm{\rho_S - \rhot_S} +
\norm{\Omega_S - \Omegat_S}  + \norm{\rhot_S - \Omegat_S}.
\end{equation}

We bound $\norm{\rho_S - \rhot_S}$ as follows:
\begin{eqnarray}
\norm{\rho_S - \rhot_S} &\leq&
\norm{\proj{\phi}-\proj{\tilde{\phi}}} \nonumber \\
&\leq& \sqrt{2} \normtwo{\proj{\phi}-\proj{\tilde{\phi}}} \nonumber \\
&=& \sqrt{2 \Tr(\proj{\phi}-\proj{\tilde{\phi}})^2} \nonumber \\
&=& \sqrt{2 (1 - 2\bra{\phi}\sqrt{X_R}\ket{\phi}^2 +
\bra{\phi}X_R\ket{\phi}^2)} \nonumber \\
&\leq& \sqrt{2 (1 - \bra{\phi}X_R\ket{\phi}^2)} \nonumber \\
&\leq& \sqrt{4 (1 - \Tr(X_R \proj{\phi}))},
\end{eqnarray}
where in the first inequality we have used the non-increase of the
trace-norm under partial tracing, in the second inequality we have
used Lemma \ref{norm_lemma} (Appendix \ref{norm_app}) and the fact
that $\ket{\phi}$ and $\ket{\tilde{\phi}}$ span a two-dimensional
subspace, and in the third inequality we have used the fact that
$X_R \leq \sqrt{X_R}$ (because $X_R \leq \identity_R$).

It follows that
\begin{eqnarray}
\av{\norm{\rho_S - \rhot_S}} &\leq& \av{\sqrt{4 (1 - \Tr(X_R
\proj{\phi}))}}\nonumber  \\
&\leq& \sqrt{ \av{ 4(1 - \Tr(X_R \proj{\phi}))}} \nonumber \\
&=& \sqrt{ 4(1 - \Tr(X_R \mathcal{E}_R )} \nonumber \\
&\leq& 2\sqrt{\delta},
\end{eqnarray}
where we have used the concavity of the square root function and
equation (\ref{delta_eqn}).

In addition, note that from the triangle inequality,
\begin{eqnarray}
\norm{\Omega_S-\Omegat_S} &=& \norm{\av{ \rho_S-\rhot_S}} \nonumber \\
&\leq& \av{\norm{\rho_S-\rhot_S}} \nonumber \\
&\leq& 2\sqrt{\delta}.
\end{eqnarray}

Inserting these results into the average of equation
(\ref{av_sum_eqn}) we get
\begin{equation}
\av{\norm{\rho_S - \Omega_S}} \leq
\sqrt{\frac{\tilde{d}_S}{\tilde{d}_E^{\eff}}} + 4 \sqrt{ \delta}
\end{equation}
and inserting this in equation (\ref{pre_calc_eqn2}) we obtain
Theorem \ref{thm:main-approximate}.

\section{Method II:  Applying Levy's \protect\\ Lemma to expectation values}
\label{sec:method-II} In this section, we describe an alternative
method of obtaining bounds on $\norm{\rho_S - \Omega_S}$ by
considering the expectation values of a complete set of
observables. The physical intuition is that if the expectations of
all observables on two states are close to each other, then the
states themselves must be close.

We begin by showing that for an arbitrary (bounded) observable
$O_S$ on $S$, the difference in expectation value between a
randomly chosen state $\rho_S =\Tr_E (\proj{\phi})$ and the
canonical state $\Omega_S$ is small with high probability. We then
proceed to show that this holds for a full operator basis, and
thereby prove that $\rho_S \approx \Omega_S$ with high probability
when $d_R \gg d_S^2$.

In this method, Levy's Lemma plays a far more central role. This
approach may be more suitable in some situations, and yields
further insights into the underlying structure of the problem.

\subsection{Similarity of expectation values for random and
canonical states} Consider Levy's Lemma applied to the expectation
value of an operator $O_S$ on $\mathcal{H}_S$, for which we take
\begin{equation}
f(\phi) = \Tr(O_S \,\rho_S).
\end{equation}
in (\ref{levy_eqn}). Let $O_S$ have bounded operator norm
$\opnorm{O_S}$ (where $\opnorm{O_S}$ is the modulus of the maximum
eigenvalue of the operator). Then the Lipschitz constant of
$f(\phi)$ is also bounded, satisfying $\eta \leq 2 \opnorm{O_S}$
(as shown in appendix \ref{lipschitz_app}). We therefore obtain
\begin{equation}
{\rm Prob}\big[ \left|\Tr(O_S \rho_S)- \langle \Tr(O_S \rho_S)
\rangle \right| \geq \epsilon \big] \leq 2 \exp \left(
\textstyle{- \frac{C d_R\, \epsilon^2}{ \opnorm{O_S}^2 }}\right).
\end{equation}

However, note that
\begin{equation}
\langle \Tr(O_S \rho_S) \rangle =  \Tr(O_S \langle\rho_S\rangle) =
\Tr (O_S \Omega_S),
\end{equation}
and hence that
\begin{equation} \label{bounded_obs_eqn}
{ \rm Prob}\big[\left|\Tr(O_S \rho_S) - \Tr (O_S \Omega_S) \right|
\geq \epsilon \big] \leq 2 \exp\left({\textstyle - \frac{C d_R
\epsilon^2}{ \opnorm{O_S}^2}}\right).
\end{equation}
By choosing $\epsilon=d_R^{-1/3}$ we obtain the result that
\begin{equation}\begin{split}
{\rm Prob} &\big[ \left|\Tr(O_S \rho_S)\!-\! \Tr(O_S \Omega_S)
\right| \geq d_R^{-1/3} \big] \\
           &\phantom{===========}
            \leq 2 \exp\left({\textstyle- \frac{ C d_R^{1/3}}{ \opnorm{O_S}^2}}\right).
\end{split}\end{equation}
For $d_R \gg 1$, the expectation value of any given bounded
operator for a randomly chosen state will therefore be close to
that of the canonical state $\Omega_S$ with high probability
\cite{expectation}.

\subsection{Similarity of expectation values for a complete operator basis}

Here we consider a complete basis of operators for the system.
Rather than Hermitian operators, we find it convenient to consider
a basis of unitary operators $U_S^{x}$. We show that with high
probability \emph{all} of these operators will have (complex)
expectation values close to those of the canonical state.

It is always possible to define $d_S^2$ unitary operators
$U_S^{x}$ on the system, labelled by $x \in \{0,1, \ldots
d_S^2-1\}$, such that these operators form a complete orthogonal
operator basis for $\mathcal{H}_S$ satisfying \cite{pauli_op}
\begin{equation}
\Tr(U_S^{x \dag} U_S^{y}) = d_S\,\delta_{xy},
\end{equation}
where $\delta_{xy}$ is the Kronecker delta function. One possible
choice of $U_S^{x}$ is given by
\begin{equation}
U_S^x = \sum_{s=0}^{d_S-1}  e^{2\pi i s (x-(x \,\textrm{mod}\,
d_S))/d_S^2} \ket{(s+x)\, \textrm{mod}\, d_S}\bra{s}.
\end{equation}
Noting that $\opnorm{U_S^x}=1 \; \forall \, x$ (due to unitarity),
we can then apply equation (\ref{bounded_obs_eqn}) to $O_S =U_S^x$
to obtain
\begin{equation}
\textrm{Prob}\big[\left|\Tr(U_S^x \rho_S) - \Tr (U_S^x \Omega_S)
\right| \geq \epsilon \big] \leq 2 e^{- C d_R \epsilon^2} \quad
\forall\, x.
\end{equation}
Furthermore, as there are only $d_S^2$ possible values of $x$,
this implies that
\begin{equation} \label{family_eqn}
\textrm{Prob}\big[ \exists \mathbf{x} :  \left|\Tr(U_S^x \rho_S) -
\Tr (U_S^x \Omega_S) \right| \geq \epsilon \big] \leq 2 d_S^2 e^{-
C d_R \epsilon^2}
\end{equation}

If we take $\epsilon=d_R^{-1/3} \ll 1$, note that as the right
hand side of (\ref{family_eqn}) will be dominated by the
exponential decay $e^{- C d_R^{1/3}}$, it is very likely that all
operators $U_S^x$ will have expectation values close to their
canonical values.

\subsection{Obtaining a probabilistic bound on $\norm{\rho_S -
\Omega_S}$}

As the $U_S^{x}$ form a complete basis, we can expand any state
$\rho_S$ as
\begin{equation}
\rho_S = \frac{1}{d_S} \sum_x C_x(\rho_S) U_S^{x}
\end{equation}
where
\begin{equation}
C_x (\rho) = \Tr( U_S^{x\dag} \, \rho_S ) = \Tr( U_S^x \, \rho_S
)^{\star}.
\end{equation}

Expressing equation (\ref{family_eqn}) in these terms we obtain
\begin{equation} \label{family_eqn2}
\textrm{Prob}\big[ \exists \,x :  \left|C_x(\rho) - C_x (\Omega)
\right| \geq \epsilon \big] \leq 2 d_S^2 e^{- C d_R\, \epsilon^2}
\end{equation}

When $|C_x (\rho_S) - C_x (\Omega_S) | \leq \epsilon$ for all $x$,
an upper bound can be obtained for the squared Hilbert-Schmidt
norm \cite{last-footnote} as follows:
\begin{eqnarray}
\normtwo{\rho_A - \Omega_A}^2 &=& \normtwo{ \frac{1}{d_S} \sum_x
\left( C_x (\rho_S) - C_x (\Omega_S) \right) U_S^{x}}^2 \nonumber
\\ \nonumber &=& \frac{1}{d_S^2} \Tr \left( \sum_x \left(
C_x (\rho_S) - C_x (\Omega_S) \right) U_S^x \right)^2
 \\ \nonumber &=& \frac{1}{d_S}
\sum_x \left( C_x (\rho_S) - C_x (\Omega_S) \right)^2
\\ &\leq& d_S \epsilon^2
\end{eqnarray}
Hence using the relation between the trace-norm and
Hilbert-Schmidt-norm (proved in appendix \ref{norm_app}),
\begin{equation}
\norm{\rho_S - \Omega_S} \leq \sqrt{d_S} \normtwo{\rho_S -
\Omega_S} \leq d_S \epsilon.
\end{equation}
Incorporating this result into equation (\ref{family_eqn2}) yields
\begin{equation} \label{before_alpha_eqn}
{\rm Prob}\big[ \norm{\rho_S-\Omega_S} \geq d_S \epsilon \big]
\leq 2 d_S^2 e^{- C d_R \epsilon^2 }.
\end{equation}

If we choose
\begin{equation}
\epsilon = \left(\frac{d_S}{d_R}\right)^{1/3}
\end{equation}
we obtain the final result that
\begin{equation} \label{after_alpha_eqn}
{\rm Prob}\big[ \norm{\rho_S-\Omega_S} \geq \frac{1}{\beta} \big]
\leq 2 d_S^2 e^{- C \beta  }.
\end{equation}
where
\begin{equation}
\beta = \left(\frac{d_R}{d_S^2}\right)^{1/3}
\end{equation}

Note that $\norm{\rho_S-\Omega_S}\approx 0$ with high probability
whenever $\beta \gg \log_2(d_S) \gg 1$, and hence when $d_R \gg
d_S^2$. This result is qualitatively similar to the result
obtained using the previous method, although it can be shown that
the bound obtained is actually slightly weaker in this case.

\section{Example: Spin chain \protect\\ with $n p$ excitations}
\label{spin_sec}

As a concrete example of the above formalism, consider a chain of
$n$ spin-1/2 systems in an external magnetic field in the $+z$
direction, where the first $k$ spins form the system, and the
remaining $n-k$ spins form the environment.  We therefore consider
a Hamiltonian of the form
\begin{equation}
H =  - \sum_{i=1}^n \frac{B}{2} \, \sigma^{(i)}_z
\end{equation}
where $B$ is a constant energy (proportional to the external field
strength), and $ \sigma^{(i)}_z$ is a Pauli spin operator for the
$i^{\mathrm{th}}$ spin.

Under these circumstances, the global energy eigenstates can be
divided into orthogonal subspaces dependent on the total number of
spins aligned with the field. We consider a restriction to one of
these degenerate subspaces $\mathcal{H}_R \in \mathcal{H}_S
\otimes \mathcal{H}_E$ in which $n p$ spins are in the excited
state $\ket{1}$ (opposite to the field) and the remaining $n(1-p)$
spins are in the ground state $\ket{0}$ (aligned with the field).

With this setup, $d_S=2^k$ and
\begin{equation}
d_R = \nCr{n}{n p}.
\end{equation}
Approximating this binomial coefficient by an exponential (as in
Appendix \ref{binomial_app}), gives
\begin{equation} \label{dr_eqn}
d_R \geq \frac{2^{n H(p)}}{n+1}
\end{equation}
where $H(p)=-p \log_2 (p) - (1-p) \log_2 (1-p)$ (the Shannon
entropy of a single spin).

From Theorem \ref{thm:main},
\begin{equation}
\mathrm{Prob}\big[ \norm{\rho_S-\Omega_S} \geq \eta \big] \leq
\eta',
\end{equation}
where
\begin{eqnarray}
\eta  &=& \epsilon  + \sqrt{ \frac{d_S}{d_E^{\eff}}}, \\
\eta' &=& 2 \exp \left( - C d_R \epsilon^2 \right).
\end{eqnarray}
In addition,
\begin{equation}
\sqrt{\frac{d_S}{d_E^{\eff}}} \leq  \sqrt{\frac{d_S^2}{d_R}} \leq
\sqrt{(n+1)}\; 2^{-(n H(p) -2 k)/2}.
\end{equation}
For an appropriate choice of $\epsilon$ (e.g. $\epsilon=d_R^{-1/3}
\ll 1$), we will obtain $\norm{\rho_S - \Omega_S} \approx 0$ with
high probability whenever
\begin{equation}
\sqrt{(n+1)}\; 2^{-(n H(p) -2 k)/2} \ll 1
\end{equation}
For fixed $p$, this condition will be satisfied for all
sufficiently large $n \gg k$.

We emphasise that our results concern the distance between
$\rho_S$ and $\Omega_S$. Computing the precise form of $\Omega_S$
is a standard exercise in statistical mechanics, which we sketch
here for completeness.

In the regime where $n\gg k^2$, the canonical state $\Omega_S$
will take the approximate form
\begin{eqnarray}
\Omega_S &=& \sum_s
\frac{(n-k)!}{d_S (np-|s|)!(n(1-p)-(k-|s|))!}\, \ket{s}\bra{s} \nonumber \\
&\approx& \sum_s  \frac{n! (np)^{|s|} (n(1-p))^{k-|s|}}{d_S \, n^k
(np)!(n(1-p))!} \, \ket{s}\bra{s} \nonumber
\\ &=& \sum_s p^{|s|} (1-p)^{k-|s|} \, \ket{s}\bra{s}
\label{omega_sum_eqn}
\\ &=& (p \ket{1}\bra{1} + (1-p) \ket{0}\bra{0})^{\otimes k}.
\end{eqnarray}
and hence the canonical state of the system will approximate that
of $k$ uncorrelated spins, each with a probability $p$ of being
excited, as expected.

To connect our result to the standard statistical mechanical
formula,
\begin{equation}
\Omega_S \propto \exp\left(-\frac{H_S}{\mathrm{k}_B T}\right)
\end{equation}
we use Boltzmann's formula relating the entropy of the environment
$S_E(|e|)$ to the number of states $N_E(|e|)$ of the environment
with a given number of excitations $|e|$ to get
\begin{eqnarray}
S_E(|e|) &=& \mathrm{k}_B \ln N_E(|e|) \nonumber \\
&=&  \mathrm{k}_B \ln \nCr{n-k}{|e|} \nonumber \\
&\approx & \mathrm{k}_B \Big( (n-k)\ln (n-k) - |e|\ln |e| \nonumber \\
&& \quad\; - (n-k-|e|)\ln (n-k-|e|) \Big),
\end{eqnarray}
where in the third line we have used Stirling's approximation.
Defining the temperature in the usual way, and noting that the
energy of the environment is given by $E=|e|B-(n-k)B/2$, we obtain
\begin{eqnarray}
\frac{1}{T} &=& \frac{\mathrm{d}S_E(E)}{\mathrm{d}E}\bigg|_{E=\av{E}} \nonumber \\
&=& \frac{1}{B} \frac{\mathrm{d}S_E(|e|)}{\mathrm{d}|e|}\bigg|_{|e|=(n-k)p} \nonumber \\
            &\approx& \frac{\mathrm{k}_B}{B} \ln\left(\frac{n-k-|e|}{|e|}\right)\bigg|_{|e|=(n-k)p}\nonumber \\
            &=& \frac{\mathrm{k}_B}{B} \ln\left(\frac{1-p}{p}\right)
            \label{T_eqn}
\end{eqnarray}

This formula expresses how the probability $p$ defines a
temperature $T$ of the environment. Rearranging equation
(\ref{omega_sum_eqn}) to incorporate equation (\ref{T_eqn}) gives
the usual statistical mechanical result
\begin{eqnarray}
\Omega_S &\approx& (1-p)^k \sum_s \left(\frac{p}{1-p}\right)^{|s|}
\, \ket{s}\bra{s} \nonumber \\
&=& (1-p)^k \sum_s \exp\left( - |s|
\ln\left(\frac{1-p}{p}\right)\right)\, \ket{s}\bra{s} \nonumber \\
&=& (1-p)^k \sum_s \exp\left(- \frac{|s| B}{\mathrm{k}_B T}\right)
\ket{s}\bra{s} \nonumber \\ &\propto& \exp\left(-
\frac{H_S}{\mathrm{k}_B T} \right).
\end{eqnarray}

\subsection{Projection on the typical subspace}
\label{typical_sub_sec}

We can obtain an improved bound on $\norm{\rho_S-\Omega_S}$ by
noting that the system state almost always lies in a typical
subspace with approximately $k p$ excitations. We make use of this
observation by applying Theorem \ref{thm:main-approximate} with a
measurement operator $X_R$ given by
\begin{equation}
X_R= \Pi_S \otimes \identity_E
\end{equation}
where $\Pi_S$ is a projector onto the typical subspace of the
system, in which it contains a number of excitations $|s|$ in the
range
\begin{equation}
k p - \xi \leq |s| \leq k p + \xi.
\end{equation}

It is easy to show, using classical probabilistic arguments (see
Appendix \ref{typical_subspace_app}), that
\begin{equation}
\Tr_R(X_R \mathcal{E}_R) = \Tr_S(\Pi_S \Omega_S) \geq 1- \delta
\end{equation}
where
\begin{equation} \label{ex_delta_eqn}
\delta = 2 \exp \left( - \frac{ \xi^2 }{4 k p (1-p)} \right)
\end{equation}

Furthermore, the dimension $\tilde{d}_S$ of the support of $X_R$
on $\mathcal{H}_S$ (which here is simply the dimension of the
typical subspace) is shown in appendix \ref{binomial_app} to be
given by
\begin{eqnarray} \label{ds_eqn}
\tilde{d}_S &=& \sum_{|s| = kp-\xi}^{kp+\xi} \nCr{k}{|s|}
\nonumber \\ &\leq& (2 \xi +1)\, 2^{k H(p) + \xi G(p)}
\end{eqnarray}
where
\begin{equation}
G(p) =\left|\frac{d H(p)}{d p} \right|= \left|
\log_2\left(\frac{p}{1-p}\right) \right|.
\end{equation}

From Theorem \ref{thm:main-approximate} we obtain:
\begin{equation}
    \mathrm{Prob}\big[ \norm{\rho_S-\Omega_S} \geq \tilde{\eta} \big] \leq \tilde{\eta}',
\end{equation}
where, using $\tilde{d}_E^{\eff} \geq d_R/\tilde{d}_S$, and
inserting the results of equations (\ref{dr_eqn}),
(\ref{ex_delta_eqn}) and (\ref{ds_eqn}),
\begin{eqnarray}
    \tilde{\eta}  &=&  \epsilon + \sqrt{(n+1)}(2\xi+1) \, 2^{(k-n/2)H(p) + \xi
    G(p)} \\
                    && \quad + \sqrt{32} \exp \left( - \frac{ \xi^2 }{8 k p (1-p)} \right) \nonumber \\
    \tilde{\eta}' &=& 2 \exp \left( - C  d_R \epsilon^2 \right).
  \end{eqnarray}

Choosing $\xi = k^{2/3}$ and $\epsilon = d_R^{-1/3}$ yields
\begin{eqnarray}
    \tilde{\eta}  &=&  (n+1)^{1/3}\, 2^{-n H(p)/3} \nonumber \\
    && \quad + \sqrt{(n+1)}(2 k^{2/3}+1) \, 2^{-(n - 2k)H(p)/2 + k^{2/3} G(p)} \nonumber \\
                    && \qquad + \sqrt{32} \exp \left( - \frac{ k^{1/3} }{8 p (1-p)} \right) \\
    \tilde{\eta}' &=& 2 \exp \left( - \frac{ C 2^{n H(p)/3}}{(n+1)^{1/3}}  \right).
\end{eqnarray}

In the thermodynamic limit in which $p$ is fixed (corresponding to
the temperature), the ratio of the system and environment sizes
$r=k/(n-k)$ is fixed at some value $r < 1$ (i.e. the system is
smaller than the environment), and $n$ tends to infinity, $\eta
\rightarrow 0$  and $\eta' \rightarrow 0$, and hence $\rho_S
\rightarrow \Omega_S$.

For large (but finite) $n$ the system will be thermalised for
almost all states when the system is smaller than the environment
(i.e. $r < 1$). Note that as $\eta$ depends exponentially on
$(n-2k)$, $\eta \ll 1$ can be achieved with only small differences
in the number of spins in the system and environment.

\section{Conclusions}
\label{sec:conclusions}

Let us look back at what we have done. Concerning the problem of
thermalisation of a system interacting with an environment in
statistical mechanics, there are several standard approaches. One
way of looking at it is to say that the only thing we know about
the state of the universe is a global constraint such as its total
energy. Thus the way to proceed is to take a Bayesian point of
view and consider all states consistent with this global
constraint to be equally probable. The average over all these
states indeed leads to the state of any small subsystem being
canonical. But the question then arises what is the meaning of
this average, when we deal with just one state. Also, these
probabilities are subjective, and this raises the problem of how
to argue for an objective meaning of the entropy. A formal way out
is that suggested by Gibbs, to consider an ensemble of universes,
but of course this doesn't solve the puzzle, because there is
usually only one actual universe. Alternatively, it was suggested
that the state of the universe, as it evolves in time, can reach
any of the states that are consistent with the global constraint.
Thus if we look at time averages, they are the same as the average
that results from considering each state of the universe to be
equally probable. To make sense of this image one needs
assumptions of ergodicity, to ensure that the universe explores
all the available space equally, and of course this doesn't solve
the problem of what the state of the subsystem is at a given time.

What we showed here is that these averages are not necessary.
Rather, (almost) any individual state of the universe is such that
any sufficiently small subsystem behaves as if the universe were
in the equiprobable average state. This is due to massive
entanglement between the subsystem and the rest of the universe,
which is a generic feature of the vast majority of states. To
obtain this result, we have have introduced measures of the
effective size of the system, $d_S$, and its environment (i.e. the
rest of the universe), $d_E^{\eff}$, and showed that the average
distance between the individual reduced states and the canonical
state is directly related to $d_S/d_E^{\eff}$. Levy's Lemma is
then invoked to conclude that all but an exponentially small
fraction of all states are close to the canonical state.

In conclusion, the main message of our paper is that averages are
not needed in order to justify the canonical state of a system in
contact with the rest of the universe -- almost any individual
state of the universe is enough to lead to the canonical state. In
effect, we propose to replace the Postulate of Equal a priori
Probabilities by the Principle of Apparently Equal a priori
Probabilities, which states that as far as the system is concerned
every single state of the universe seems similar to the average.

We stress once more that we are concerned only with the distance
between the state of the system and the canonical state, and not
with the precise mathematical form of this canonical state.
Indeed, it is an advantage of our method that these two issues are
completely separated. For example, our result is independent of
the canonical state having Boltzmannian form, of degeneracies of
energy levels, of interaction strength, or of energy (of system,
environment or the universe) at all.

In future work \cite{dynamics}, we will go beyond the kinematic
viewpoint presented here to address the dynamics of
thermalisation. In particular, we will investigate under what
conditions the state of the universe will evolve into (and spend
almost all of its later time in) the large region of its Hilbert
space in which its subsystems are thermalised.

\acknowledgments The authors would like to thank Yakir Aharonov
and Noah Linden for illuminating discussions. SP, AJS and AW
acknowledge support through the U.K.~EPSRC's project ``QIP IRC''.
In addition, SP also acknowledges support through EPSRC
``Engineering-Physics'' grant GR/527405/01, and AW acknowledges a
University of Bristol Research Fellowship.

\bigskip \noindent \textbf{Note added}: A very recent independent paper by
Goldstein et. al. \cite{goldstein} discusses similar issues to
those addressed here.


\appendix

\section{Lipschitz constants \protect\\ and norm relation}
\label{lipschitz_app}\label{norm_app}
\begin{lemma}
The Lipschitz constant $\eta$ of the function
$f(\phi)=\norm{\rho_S-\Omega_S}$, satisfies $\eta \leq 2$.
\end{lemma}
\textbf{Proof: } Defining the reduced states
$\rho_{1}=\Tr_E(\proj{\phi_{1}})$ and
$\rho_2=\Tr_E(\proj{\phi_{2}})$, and using the result that partial
tracing cannot increase the trace-norm
\begin{eqnarray}
\left| f(\phi_1) - f(\phi_2) \right|^2 &=& \left| \norm{\rho_1 -
\Omega} - \norm{\rho_2 - \Omega}
\right|^2 \nonumber \\
&\leq& \norm{\rho_1 - \rho_2}^2 \nonumber  \\
&\leq&  \norm{\proj{\phi_1} - \proj{\phi_2}}^2 \nonumber  \\
&=& 4 \left( 1- \left| \langle \phi_1 | \phi_2 \rangle
\right|^2 \right) \nonumber \\
&\leq& 4 \left| \ket{\phi_1} - \ket{\phi_2} \right|^2
\end{eqnarray}
Hence $\left| f(\phi_1) - f(\phi_2) \right| \leq 2 \left|
\ket{\phi_1} - \ket{\phi_2} \right|$, and thus $\eta \leq
2$.$\square$

\begin{lemma}
The Lipschitz constant $\eta$ of the function $f(\phi) = \Tr( X
\proj{\phi})$, where $X$ is any operator on $\mathcal{H}_R$ with
finite operator norm $\opnorm{X}$ satisfies $\eta \leq
2\opnorm{X}$.
\end{lemma}
\textbf{Proof: }
\begin{eqnarray}
\left| f(\phi_1) - f(\phi_2) \right| &=& \big| \bra{\phi_1} X
\ket{\phi_1} -\bra{\phi_2} X \ket{\phi_2} \big|  \nonumber \\
\nonumber &=& \frac{1}{2} \big| (\bra{\phi_1} + \bra{\phi_2} ) X (
\ket{\phi_1} - \ket{\phi_2} )  \nonumber \\ && \quad  +
(\bra{\phi_1}
- \bra{\phi_2} ) X ( \ket{\phi_1} + \ket{\phi_2} )  \big| \nonumber \\
&\leq& \opnorm{X} \big|\ket{\phi_1} + \ket{\phi_2}\big|
\big|\ket{\phi_1} - \ket{\phi_2}\big| \nonumber  \\
&\leq& 2\opnorm{X} \big|\ket{\phi_1} - \ket{\phi_2}\big|. \quad
\square
\end{eqnarray}

\begin{lemma} \label{norm_lemma}
For any $n\times n$ matrix $M$,  $\norm{M} \leq
\sqrt{n} \normtwo{M}$.
\end{lemma}
\textbf{Proof: } If $M$ has eigenvalues $\lambda_i$,
\begin{equation*}\begin{split}
  \norm{M}^2 &=    n^2 \left( \frac{1}{n} \sum_i |\lambda_i| \right)^2 \\
             &\leq n^2 \frac{1}{n}\sum_i |\lambda_i|^2
              =    n\normtwo{M}^2,
\end{split}\end{equation*}
by the convexity of the square function.
Taking the square-root yields the desired result. $\square$

\section{Projection onto \protect\\ the typical subspace}
\label{typical_subspace_app}

\begin{lemma} \label{typ_lemma}
Given a system in the canonical state $\Omega_S$, the probability
of it containing a number of excitations $|s|$ in the range $kp -
\xi \leq |s| \leq kp +\xi$ is given by
\begin{equation}
\Tr (\Pi_S \Omega_S) \geq 1- \delta
\end{equation}
where
\begin{equation}
\delta = 2 \exp \left( - \frac{ \xi^2 }{4 k p (1-p)} \right)
\end{equation}
\end{lemma}
\textbf{Proof:}  $\Omega_S$ is essentially a classical
probabilistic state, obtained by choosing $k$ spins at random from
a `bag' containing $n p$ excited spins and $n(1-p)$ un-excited
spins without replacement. It is easy to see that this state will
lie in the typical subspace with higher probability than if the
spins were replaced in the bag after each selection, as the former
process is mean reverting, whereas the latter is not. We can bound
the probability of lying outside the typical subspace in the case
with replacement using Chernoff's inequality \cite{chernoff} for
the sum $X=\sum_i (s_i-p)$, where $s_i\in\{0,1\}$ is the value of
the $i^{\mathrm{th}}$ spin. This gives
\begin{equation}
\textrm{Prob}\Big[\big| X \big| > \xi \Big] \leq 2
e^{-\frac{\xi^2}{4 \sigma^2}}
\end{equation}
where $\sigma^2 = k p (1-p)$ is the variance of $X$. Hence
\begin{equation}
\textrm{Prob}\Big[\big| |s| - k p  \big| > \xi \Big] \leq 2
e^{-\frac{\xi^2}{4 k p (1-p)}}
\end{equation}
and thus
\begin{eqnarray}
\Tr (\Pi_S \Omega_S ) &=& 1 - \textrm{Prob}\Big[\big| |s| - k p
\big| > \xi \Big] \nonumber \\
                    &\geq& 1 - 2 e^{-\frac{\xi^2}{4 k p (1-p)}} \quad
                    \square
\end{eqnarray}

\section{Exponential bounds \protect\\ on combinatorial quantities}
\label{binomial_app}
In this appendix we obtain bounds for the combinatoric quantities
required to consider the example case of a spin-chain.

From standard probability theory we know that
\begin{equation}
\sum_{k=0}^{n} \nCr{n}{k} p^k (1-p)^{n-k} = 1,
\end{equation}
with the maximal term in the sum being obtained when $k=np$. Hence
\begin{equation} \label{app_eqn1}
\nCr{n}{np}\! p^{np} (1-p)^{n(1-p)} \leq 1 \leq \sum_{k=0}^n\!
\nCr{n}{np}\! p^{np} (1-p)^{n(1-p)} .
\end{equation}
Noting that
\begin{equation}
p^{np} (1-p)^{n(1-p)} = 2^{-n H(p)}
\end{equation}
where $H(p)=-p \log_2 (p) - (1-p) \log_2 (1-p)$, we can rearrange
equation (\ref{app_eqn1}) to get
\begin{equation}
2^{n H(p) - \log_2 (n+1)} \leq \nCr{n}{np} \leq 2^{n H(p)}.
\end{equation}

We also require an upper bound for the dimension of the typical
subspace of system $S$, given by
\begin{equation}
  \tilde{d}_S = \sum_{|s| = kp- \xi}^{kp+\xi} \nCr{k}{|s|}.
\end{equation}
The maximal term in this sum occurs when $|s|=k \tilde{p}$ where
\begin{equation}
\tilde{p} = \left\{ \begin{array}{lcl} p+\xi/k &:&
p<\frac{1}{2}-\xi/k
\\\frac{1}{2} &:& \left|p- \frac{1}{2}\right| \leq \xi/k
\\ p-\xi/k &:&
p>\frac{1}{2}+\xi/k, \end{array} \right.
\end{equation}
and as the sum consists of $(2 \xi + 1)$ terms,
\begin{equation}
 \tilde{d}_S \leq (2 \xi + 1) \nCr{k}{k
\tilde{p}}.
\end{equation}
Bounding the binomial coefficient by an exponential as above we
obtain
\begin{equation}
\tilde{d}_S \leq (2 \xi + 1) 2^{k H(\tilde{p})}.
\end{equation}
As $H(p)$ is a concave function of $p$, we also note that
\begin{equation}
k H(\tilde{p}) - k H(p) \leq  \xi \left|\frac{d H(p)}{d p} \right|
\end{equation}
Defining
\begin{equation}
G(p) =\left|\frac{d H(p)}{d p} \right|= \left|
\log_2\left(\frac{p}{1-p}\right) \right|,
\end{equation}
we therefore find that
\begin{equation}
\tilde{d}_S \leq (2 \xi + 1) 2^{k H(p) + \xi \,G(p)}.
\end{equation}


\begin{thebibliography}{99}


\bibitem{aharonov} Y. Aharonov, personal communication, 1986.


\bibitem{mahler} J. Gemmer, M. Michel, and G. Mahler,
\emph{Quantum Thermodynamics}, LNP 657, Springer Verlag, Heidelberg, Berlin, 2004.

\bibitem{levy1} V. D. Milman and G. Schechtman,
\emph{Asymptotic Theory of Finite-Dimensional Normed Spaces}, LNM
1200, Appendix IV, Springer Verlag, 1986.

\bibitem{levy2} M. Ledoux, \emph{The concentration of measure
phenomenon}, AMS Mathematical Surveys and Monographs, vol. 89,
American Mathematical Society, 2001.

\bibitem{hayden} P. Hayden, D. W. Leung, and A. Winter, eprint
 {\tt quant-} {\tt ph/0407049} (2004). To appear in Commun. Math. Phys.

\bibitem{trace-dist} Note that $\norm{\rho_S-\Omega_S}$ is twice the
usual trace-distance between the two states, taking a maximum
value of 2 for orthogonal states.


\bibitem{smooth-renyi} R. Renner and R. K\"onig, Proc. TCC 2005, LNCS 3378, Springer
Verlag, 2005. R. Renner and S. Wolf, Proc. 2004 IEEE Intl. Symp. Inf. Theory, p.~233
(2004).

\bibitem{hayden:winter} P. Hayden and A. Winter, Phys. Rev. A {\bf 67},
012326 (2003), eprint {\tt quant-ph/0204092} (2002).

\bibitem{lloyd-shor} S. Lloyd, Phys. Rev. A {\bf 55}, 1613 (1997).
P. W. Shor, unpublished MSRI lecture notes (Berkeley, 2002);
available online at {\tt www.msri.org/}{\tt publications/ln/msri/2002/}
{\tt quantumcrypto/shor/1/}.

\bibitem{HOW} M. Horodecki, J. Oppenheim, and A. Winter,
Nature {\bf 436}, 673 (2005). Long version in preparation.

\bibitem{support} Mathematically, $\tilde{d}_S =
\min_{\Pi_S} \Tr \Pi_S$, where the minimum is over all projectors
$\Pi_S\in\mathcal{H}_S$ such that $[(\Pi_S \otimes \identity_E),
X_R]=0$.

\bibitem{fuchs:vandegraaf} C. A. Fuchs and J. van de Graaf,
IEEE Trans. Inf. Theory {\bf 45}, 1216 (1999).

\bibitem{expectation} Note that the probability of obtaining a
particular outcome in a measurement is always representable as the
expectation value of a bounded observable (with $\|O_S\| \leq 1$),
hence this is a very general result.

\bibitem{pauli_op} In the special case in which $\log_2 d_S$ is an integer, it is also
possible to make $U_S^{x}$ Hermitian by constructing them from
tensor products of 2-dimensional Pauli spin matrices and Identity
matrices.

\bibitem{last-footnote} Note that unlike in the
previous method, it is possible to proceed directly with the
trace-norm here, but the bound obtained is weaker.

\bibitem{chernoff} A. Dembo, O. Zeitouni, \emph{Large Deviations: Techniques and
Applications}, 2nd ed., Springer Verlag, New York, 1998.

\bibitem{dynamics} Y. Aharonov, N. Linden, S. Popescu, A. J. Short,
and A. Winter. In preparation.

\bibitem{goldstein} S. Goldstein, J. L. Lebowitz, R. Tumulka,
N.Zangh\`{\i}, eprint {\tt cond-mat/0511091} (2005).

\end{thebibliography}
\end{document}